\documentclass[a4paper,11pt]{article}

\pdfoutput=1 
\usepackage{jcappub}

\usepackage{graphicx}
\usepackage{longtable}
\usepackage{float}
\usepackage{dcolumn}
\usepackage{bm}
\usepackage{appendix}
\usepackage{multirow}
\usepackage{color}
\usepackage[utf8]{inputenc}
\usepackage{footmisc}
\usepackage{txfonts}
\usepackage{xcolor}
\usepackage{graphicx}
\usepackage{bm}
\usepackage{ulem}
\usepackage{multirow, array, float, tabularx}
\usepackage{amsmath}
\usepackage[colorlinks=true,linkcolor=blue,citecolor=blue]{hyperref}

\begin{document}

\title{Brane inflation driven by an arctan potential: CMB constraints and Reheating}
%

\author[a]{R. M. P. Neves}
\emailAdd{raissapimentel.ns@gmail.com}

\author[b]{S. Santos da Costa}
\emailAdd{simonycosta@on.br}

\author[a,c]{F. A. Brito}
\emailAdd{fabrito@df.ufcg.edu.br}

\author[b]{J. S. Alcaniz}
\emailAdd{alcaniz@on.br}

\affiliation[a]{Departamento de F\'isica, Universidade Federal da Para\'iba, 58051-970, Jo\~ao Pessoa, Para\'iba, Brasil}
\affiliation[b]{Observat\'orio Nacional, 20921-400, Rio de Janeiro, RJ, Brasil}
\affiliation[c]{Departamento de F\'{\i}sica, Universidade Federal de Campina Grande, 58429-900, Campina Grande, Para\'iba, Brasil}

\abstract{
    We investigate the early universe evolution in the context of brane inflation driven by a  supergravity-inspired $\arctan$ potential. We performed a slow-roll and a semi-analytical reheating analyses and obtained constraints on the inflationary parameters in agreement with Planck 2018 data. We also employed a Markov Chain Monte Carlo analysis to perform a parameter estimation of the cosmological parameters, obtaining results in good agreement with the currently available cosmic microwave background and baryon acoustic oscillation data. This work establishes the general theoretical predictions of the $\arctan$ model, with  the results of the statistical analysis corroborating its observational viability.
    
}

\maketitle

\section{Introduction}

Measurements of the temperature fluctuations of the Cosmic Microwave Background (CMB) have provided strong observational support for the inflationary scenario~ \cite{Aghanim:2015xee,Planck,Planckk} (see also \cite{ijjas,Lindee,guth14,brandenberger} for different points of view of the current observational status of inflation). Although compatible with the simplest slow-roll scenarios of inflation, and showing a preference for plateau over monomial potentials, these observations  have also severely constrained specific class of models that emerged as an attempt to explain early accelerated phase of the Universe \cite{Lindee, SantosdaCosta:2017ctv, Santos, bbq-2007,Bhattacharya:2020gnk}.

On the other hand, the aforementioned models have also been constrained from theoretical arguments in the realm of fundamental theories such as supergravity/string theories. Since these theories live in ten or eleven dimensions, an important mechanism such as compactification of extra dimensions is necessary. However, there is an additional difficulty in producing four-dimensional effective theories able to describe inflation \cite{Ohta:2003ie,gibbons,KKLT-0,townsend,KKLT}. Alternatively, it is hard to produce four-dimensional theories that can develop a de Sitter vacuum, which is crucial for the existence of an inflationary phase of our Universe. This is because the obtained inflaton potentials are normally very steep, which do not meet the essential criteria for developing sufficiently inflation. More recently this problem has been reconsidered in a similar concept now well-known as swampland conjectures (see e.g. \cite{swamp0,swamp1,swamp2,swamp3,Garg:2018reu}).

In a previous communication \cite{Brito}, some of us considered that Bogomol'nyi-Prasad-Sommerfield (BPS) solutions of truncated supergravity theories in five-dimensions can induce four-dimensional models that are not constrained by compactification because they are induced due to force along inter-brane distance. So even if inflaton potentials coming from the dimensional reduction cannot in general produce sufficiently inflation, one can still expect that such inter-brane force can produce the desired inflaton potential to describe inflation. In the setups \cite{Brito} and more recently in \cite{Neves:2020lru} the authors considered elastic collision of bulk particles with parallel domain walls ({\it thick branes}) embedded in five-dimensions. The {\it resonant tunneling effect} that affects transmission rate through the barriers related to the parallel domain walls induces an attractive force that associated with the reflection rate allows to find an attractive $\arctan$-type potential for the inflaton field.
We have indeed followed previous attempt by Dvali and Tye in the context of producing brane inflation with extra dimensions and several sources of force.  In our setup, however, we mainly consider the dominance of particle collisions and the electric force of possibly charged domain walls \cite{Neves:2020lru}. 

Our aim in this paper is twofold: first, to perform a general analysis of the theoretical predictions of a four-dimensional $\arctan$-type inflaton potential; second, to explore its cosmological consequences and observational viability in light of currently available data. The main features of the potential is discussed through
a slow-roll analysis. Considering the results of such analysis, the reheating phase is also studied through a semi-analytical approach. Finally, we also employed a Markov Chain Monte Carlo analysis to perform a parameter estimation of the cosmological parameters using the currently available cosmic microwave background and baryon acoustic oscillation data. 

We organize this work as following: Section \eqref{model} presents the theoretical motivation for the brane scenario considered in this paper that comes from a supergravity inspired model. Section \eqref{slowrollanalysis} shows the slow-roll analysis for the potential induced on the brane, where we put theoretical constraints on the inflationary parameters. The reheating phase of the model is studied in Sec. \eqref{reheating}. We present in Section \eqref{analysisandresults} the method used to estimate the cosmological parameters,  the observational data sets used in the analysis, and also the main results obtained. Finally, we summarized the main conclusions in Section \eqref{conclusions}.

\section{The Arctan model}\label{model}

To the best of our knowledge the $\arctan$ model described in this section is the simplest model that can be found with minimal assumptions in the context of brane inflation in the realm of a string/supergravity inspired theories. These fundamental theories have a lot of constraints on the inflaton potentials, with some that run from those that invoke time-varying compact hyperbolic manifold \cite{townsend} to others that take advantage of flux compactification~\cite{KKLT}. 

The potential analysed in this work was inspired by the brane inflation scenario of Ref.~\cite{Dvali} and constructed in Ref.~\cite{Neves:2020lru}. In this scenario, the universe is described as a (3+1) dimensional thick domain wall ({\it thick brane}) embedded into a five-dimensional bulk. The interaction with another parallel brane due to elastic collisions of bulk particles induces acceleration of the universe.

We consider the scalar bosonic sector of a supergravity theory in 5D with Lagrangian given by \cite{Brito, M, F, D, Bazeia}:
\begin{eqnarray}
	e^{-1}{\cal L}_{sugra}&=&-\frac{1}{4}M^3_*R_{(5)}+G_{AB}\partial_\mu\Phi^A\partial^\mu\Phi_B 
	-\frac{1}{4}G^{AB}\frac{\partial W(\Phi)}{\partial\Phi^A}\frac{\partial W(\Phi)}{\partial\Phi^B}
	+\frac{1}{3}\frac{1}{M^3_*}W(\Phi)^2,
\label{1}
\end{eqnarray}
where $G_{AB}$ is the metric on the real scalar field space and $e=|\det g_{\mu\nu}|^{1/2}$. $R_{(5)}$ is the Ricci scalar and $1/M_{*}$ represents the five-dimensional Planck length. The superpotential $W(\Phi)$ is normally constrained and so is the five-dimensional scalar potential. As such, an effective four-dimensional theory with the induced inflaton potential with sufficient flatness to produce enough inflation is hard to find. However, it is possible to find such inflaton potential by assuming a few conditions if one takes the advantage of the following mechanism. In our model we just need the BPS domain wall solutions embedded in the 5D bulk can suffer interactions, with mainly interaction due to an attractive potential induced by bulk particle collisions with the transmission coefficient 
\begin{equation}
T=\frac{4}{\left(4\theta^2+\frac{1}{4\theta^2}\right)\cos^2{L}+4\sin^2{L}},
\end{equation}
through two parallel domain walls. Here $L\equiv L(r)$ is a function of distance $r$ that separates the barriers related to the equation of small perturbations around the domain walls and $\theta$ encodes information about the height and thickness of each barrier in terms of the colliding particles energy.  Since the reflection coefficient changes because of the {\it resonant tunneling effect} that increases transmission rate as the barriers are brought close together, few bulk particles are reflected by the domain walls and then a very small force acts to them.

\begin{figure}[ht]
\begin{center}
\includegraphics[scale=0.45]{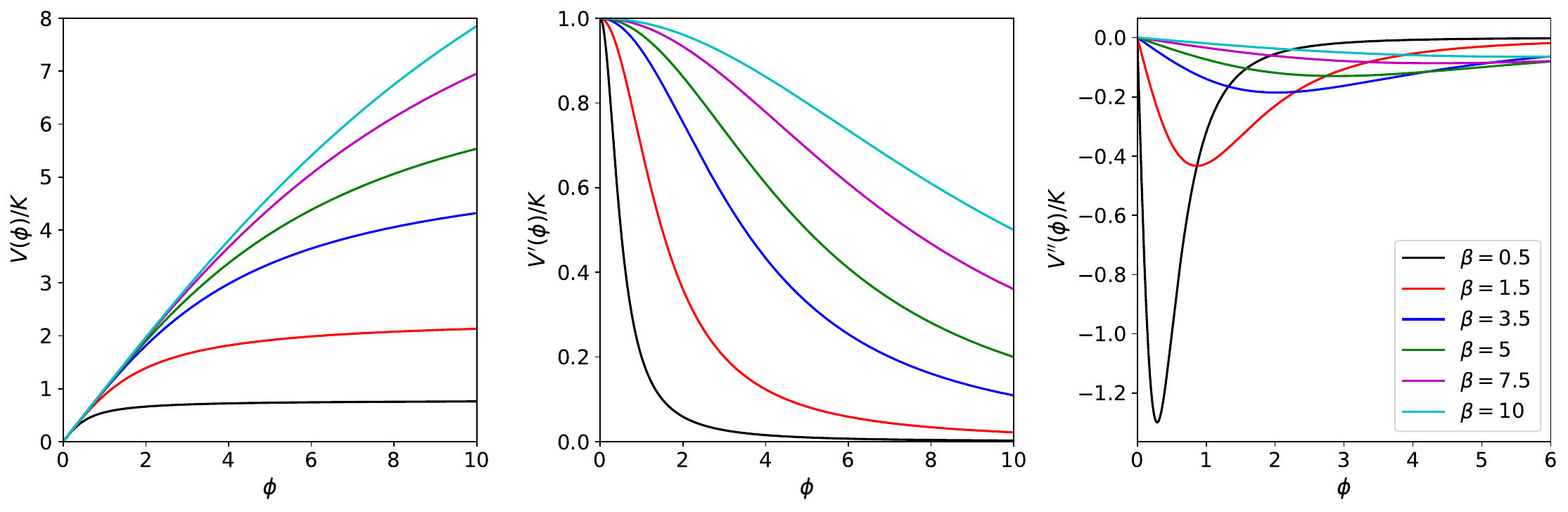}
\end{center}
\caption{The potential of the Arctan model and its first and second derivatives with respect to the field $\phi$, considering different values of $\beta$.}
\label{fig:pot_derivs}
\end{figure}

On the other hand, if one brings the domain walls far from each other the transmission rate decreases and so the reflection rate is increased and a stronger attractive force is experienced by the domain walls. By associating such force with the reflection rate it is possible to find an attractive potential as a function of the inflaton field 
\begin{eqnarray}
	V(\phi)=K\beta\arctan\left(\frac{\phi}{\beta}\right),
\label{eq:arctan_pot}
\end{eqnarray}
where the inter domain walls distance $r$ was associated with inflaton $\phi\sim \sqrt{T_{wall}}\,r$, being $T_{wall}$ the domain wall tension --- for further details see \cite{Neves:2020lru}.




We will study some theoretical and observational predictions of the potential \eqref{eq:arctan_pot},
hereafter named Arctan model, where its first and second derivatives with respect to the field $\phi$ are given by:
\begin{equation}
	V^{'}(\phi)=\frac{K\beta^{2}}{\beta^{2}+\phi^{2}} \qquad \textrm{and} \qquad V^{''}(\phi)=-\frac{2K\beta^2\phi}{\left(\phi^{2}+\beta^{2}\right)^{2}}.
\end{equation}
The potential behavior $V(\phi)$ as a function of the field $\phi$ (and its derivatives) is (are) displayed in Fig.~\eqref{fig:pot_derivs}. 
We have tried for both positive and negative values of $\beta$ and observed that the potential has the same form. This is due to the fact that the Arctan function is odd, so the multiplicative factor $\beta$ redresses the potential to behaves equally independent of the signal of $\beta$. This allows us to choose to work only with positive values of $\beta$ without losing generality.

It is worth mentioning that exists in the literature a model called Arctan, but it is different from the Arctan model we studied. The Arctan model was originally introduced in Ref.~\citep{Wang:1997cw} as a toy model where the equation of state changes rapidly around $\phi=0$. To that model, a slow-roll approximation was performed and the Bayesian evidence and complexity were calculated in light of Planck 2013 Cosmic Microwave Background data, as you can see in Refs. \citep{Martin:2013tda, Martin:2013nzq}.  

\section{Slow-roll analysis}\label{slowrollanalysis}

The slow-roll regime is characterized by the parameters $\epsilon$ and $\eta$, such that the conditions $\epsilon, \eta \ll 1$ are satisfied~\footnote{These conditions guarantee that the field slowly rolls down its potential until its minimum and that the expansion rate, $H$, is almost a constant.}. For the Arctan potential considered here, these two parameters are written in function of the potential and its first and second derivatives with respect to $\phi$ as~\cite{liddle,Lyth_1999}:
\begin{eqnarray}
	\epsilon=\frac{1}{2}\left(\frac{V'}{V}\right)^2=\frac{1}{2\beta^{2}\left[\arctan\left(\frac{\phi}{\beta}\right)\left(1+\frac{\phi^{2}}{\beta^{2}}\right)\right]^{2}}, 
\label{eq:epsilon}
\end{eqnarray}
and
\begin{equation}
	\eta=\frac{V''}{V}=-\frac{2\phi}{\beta^{3}\arctan\left(\frac{\phi}{\beta}\right)\left(1+\frac{\phi^{2}}{\beta^{2}}\right)^{2}}.
\end{equation}
When the condition $\epsilon =1$ is satisfied we can define the value of the field at the end of inflation, $\phi_{end}$. However, we could not invert Eq.~\eqref{eq:epsilon} directly, instead, we did it numerically considering the values of $\phi_{end}$ and $\beta$ which satisfy $\epsilon(\phi_{end})=1$. In order to determine a smooth function which passes exactly through these points, and which can be evaluated everywhere, we have used a routine to interpolate the data, from which we could fit a polynomial function of 15th order necessary to calculate other quantities (as we shall see).

Concerning the amplitude $K$ of the potential \eqref{eq:arctan_pot}, we consider the primordial power spectrum of curvature perturbations, determined at the scale when the CMB crosses the Hubble horizon during inflation as
 \begin{eqnarray}
  P_{R}=\frac{V(\phi)}{24\pi^{2}\epsilon}\mid_{k=k_{\ast}}.
 \label{cap6_eq_43}
 \end{eqnarray}
The value of $P_R(k_{*})$ is determined by Planck normalization to $2.0933\times 10^{-9}$ for the pivot choice $k_{*}=0.05$Mpc$^{-1}$~\citep{Aghanim:2018eyx}.
Hence, using the potential $V(\phi)$ given by Eq.~\eqref{eq:arctan_pot} and $\epsilon$ given by \eqref{eq:epsilon}, and inverting for $K$, we obtain:
\begin{eqnarray}
	K=\frac{12\pi^{2}P_{R}(k_{\ast})}{\left(1+\frac{\phi_{k}^2}{\beta^2}\right)^2\left[\beta\arctan\left(\frac{\phi_{k}}{\beta}\right)\right]^3}.
\label{amplitudeK}
\end{eqnarray}

We can find the value of the field $\phi_{k}$ using the expression for the number of e-folds, since the horizon crossing moment up to the end of inflation
\begin{eqnarray}
	{\cal N}_k=\ln{\left(\frac{a_{end}}{a_{k}}\right)}&=& \int_{\phi_{end}}^{\phi_{k}}{\frac{d\phi}{\sqrt{2\epsilon}}}\nonumber \\
	&=& \left(\frac{\phi^{3}}{3\beta}+\phi\beta\right)\arctan\left(\frac{\phi}{\beta}\right)-\frac{\phi^{2}}{6}-\frac{\beta^{2}}{3}\ln\left(1+\frac{\phi^{2}}{\beta^{2}}\right)_{\phi_{end}}^{\phi_{k}}.
	\label{eq:efolds}
\end{eqnarray}
We then consider the pivot scale for which the CMB mode crosses the Hubble horizon during inflation to be $\mathcal{N}=55$. Since the previous equation can not be solved analytically, we have used numerical methods to solve it, in order to find the values of $\phi_{k}$ and $\beta$ that satisfy $\mathcal{N}=55$. Thus, it is possible to interpolate the data and obtain a polynomial fit for $\phi_{k}$.
Similarly, in order to find the value of the field at the beginning of inflation, we solved numerically the equation for the number of e-folds, considering $\mathcal{N}=70$,  and then interpolate a polynomial fit for $\phi_{ini}$.
For both cases, $\phi_{k}$ and $\phi_{ini}$, the polynomial functions obtained were of 15th order. 
Furthermore, we can apply this method to get the field when $\mathcal{N}=50$ and $\mathcal{N}=60$, in order to calculate the predictions for the scalar spectral index, $n_s$, and the tensor-to-scalar ratio, $r$. These inflationary parameters can be written as
\begin{equation}
	n_{s}=1-6\epsilon +2\eta=1-\frac{1}{\beta^{2}\arctan\left(\frac{\phi}{\beta}\right)\left(1+\frac{\phi^{2}}{\beta^{2}}\right)^{2}}\left[\frac{4\phi}{\beta} + \frac{3}{\arctan\left(\frac{\phi}{\beta}\right)}\right],\label{eq:ns}
\end{equation}
and
\begin{equation}
	r=16\epsilon=\frac{8}{\beta^{2}\left[\arctan\left(\frac{\phi}{\beta}\right)\left(1+\frac{\phi^{2}}{\beta^{2}}\right)\right]^{2}}
\end{equation}

%
\begin{figure}[!ht]
 \begin{center}
 \includegraphics[scale=0.7]{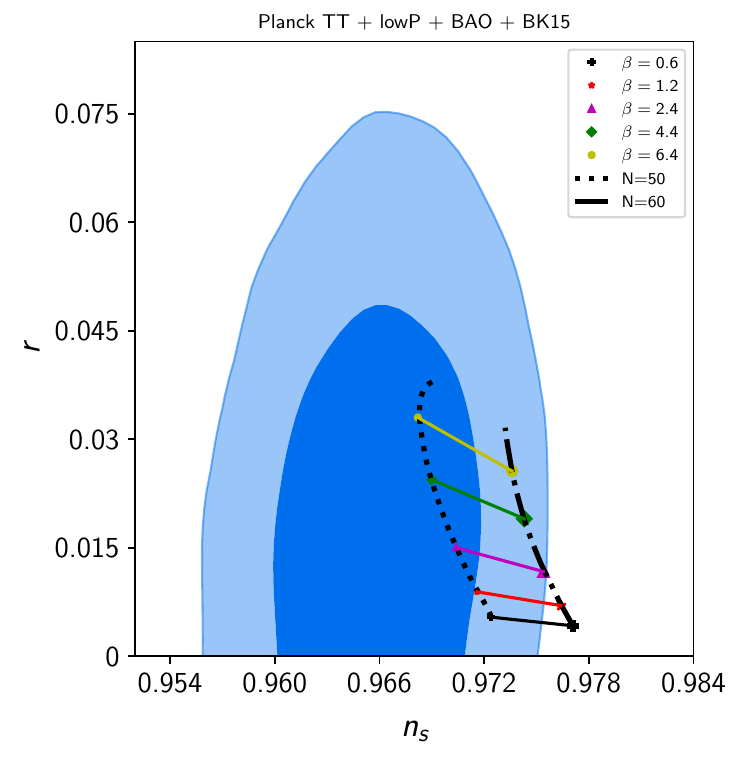}
 \end{center}
\caption{The cosmological observables $n_s$ and $r$ for the Arctan potential \eqref{eq:arctan_pot}, considering different values of the parameter $\beta$ and two values for the number of e-folds, $N=50$ and $N=60$, respectively. The contours correspond to the Planck (2018)+BAO+BICEP2/Keck data ($68\%$ and $95\%$ C.L.) using the pivot $k_\star$= 0.05 Mpc$^{-1}$.}
\label{fig:nsr_plan}
\end{figure}

We show the behavior of the spectral index and the tensor-to-scalar ratio for different values of $\beta$ in Fig.~\eqref{fig:nsr_plan}, considering the number of e-folds given in Eq.~\eqref{eq:efolds} ranging from $\mathcal{N}=50$ to $\mathcal{N}=60$.
We used the number of e-folds in this range in order to compare the model predictions to the Planck data, which are displayed as the confidence regions corresponding to $68\%$ and $95\%$, obtained from the latest release of Planck CMB temperature data~\citep{Aghanim:2018eyx} combined with Baryon Acoustic Oscillations and the tensor amplitude of B-mode polarization from the BICEP2/Keck data. Notice that the value of $n_s$ slightly decreases as $\beta$ increases while $r$ increase with the $\beta$ increases. The results for the $n_s - r$ plan exhibit good agreement with the CMB data, with almost all the values inside at least $2\sigma$ C.L., which allow us to consider as an appropriate range for $\beta$ in our analysis the interval $0.6<\beta<7$.

\section{Reheating phase after slow-roll}\label{reheating}

According to the Big Bang cosmology, the evolution history of the universe is well established as a sequence of domination eras: radiation followed by matter ending in the current accelerated phase. In addition, previously to the radiation epoch, we also had a primordial accelerated phase, i.e. inflationary phase, whose ending should allow the universe to be reheated in order to the subsequent evolution. This process, known as reheating phase, converts the energy density in the inflaton to the thermal bath, at a reheating temperature, $T_{re}$, that fills the Universe at the beginning of the standard radiation-dominated epoch. Therefore, by studying reheating we can justify the usual approach of considering the range $50<\mathcal{N}<60$ for the number of e-folds in the $n_s-r$ plane, but we may also bring back to the game that models discarded by the instantaneous reheating. 

We follow previous works~\cite{Dai:2014jja,Munoz:2014eqa,Cook:2015vqa,Ueno:2016dim} and consider that during the reheating epoch the universe is dominated by an energy component with an effective equation-of-state parameter $w_{re}$, such that its energy density decays as $\rho \propto a^{-3(1+w_{re})}$. It is worth mentioning that the physics of reheating is far away to be settled, but there is a simple canonical scenario~\cite{Abbott1982,Dolgov:1982th,Albrecht:1982mp} where the relativistic particles arise from the oscillations of the inflaton about the minimum of its potential. In this scenario, the reheating phase lasts for a time $\sim\Gamma^{-1}$, where $\Gamma$ is the inflaton-decay rate and the effective equation-of-state parameter is $w_{re}=0$. Other possibilities may also be considered, however. In fact, the bottom line is that $w_{re}>-1/3$ is needed to end inflation and numerical studies of the aforementioned thermalization has suggested a range of variation of $0<w_{re}<0.25$~\cite{Dai:2014jja,Munoz:2014eqa,Cook:2015vqa,Ueno:2016dim}, in such a way that we can consider as an adequate range  $-1/3<w_{re}<1/3$.

The number of e-folds of this epoch can be written as
\begin{eqnarray}
    \mathcal{N}_{re}=\ln{\left(\frac{a_{re}}{a_{end}}\right)}=\frac{1}{3(1+w_{re})}\ln{\left(\frac{\rho_{end}}{\rho_{re}}\right)},\label{efoldre}
\end{eqnarray}
where $a_{re}$ is the scale factor at the end of reheating. At this epoch, we can consider the energy density of the universe as
\begin{equation}
    \rho_{re}=\frac{\pi^2g_{re}}{30}T_{re}^4,
\end{equation}
where $g_{re}$ is the number of internal degrees of freedom of relativistic particles at the end of reheating, which we assume to be $g_{re}=\mathcal{O}(100)$. Using the Friedmann equations we can also demonstrate that at the end of inflation $\rho_{end}=\frac{3}{2}V_{end}$. This allows to rewrite \eqref{efoldre} as
\begin{eqnarray}
   \mathcal{N}_{re}= \frac{1}{3(1+w_{re})}\ln{\left(\frac{30\frac{3}{2}V_{end}}{\pi^2g_{re}T_{re}^4}\right)}\label{eqNre1}
\end{eqnarray}
At this point, we use the entropy conservation between the end of reheating and today to relate the temperatures in terms of the helicity states in the radiation gas, such that~\citep{Cook:2015vqa}:
\begin{eqnarray}
  T_{re}=T_0\left(\frac{a_0}{a_{re}}\right)\left(\frac{43}{11g_{re}}\right)^{\frac{1}{3}}=T_0\left(\frac{a_0}{a_{eq}}\right)e^{ \mathcal{N}_{RD}}\left(\frac{43}{11g_{re}}\right)^{\frac{1}{3}},\label{eqtemp1}
\end{eqnarray}
where $ \mathcal{N}_{RD}$ is the length in e-folds of radiation dominance, i.e. $e^{- \mathcal{N}_{RD}}\equiv{a_{re}}/{a_{eq}}$. To obtain the ratio $a_0/a_{eq}$, we can use its equivalent:
\begin{equation}
    \frac{a_0}{a_{eq}}=\frac{a_0H_k}{k}\frac{a_k}{a_{end}}\frac{a_{end}}{a_{re}}\frac{a_{re}}{a_{eq}}=\frac{a_0H_k}{k}e^{- \mathcal{N}_k}e^{- \mathcal{N}_{re}}e^{- \mathcal{N}_{RD}},
\end{equation}
where we have used the comoving Hubble scale\footnote{For the pivot choice $k=0.05$Mpc$^{-1}$ at which Planck determines $n_s$~\cite{Aghanim:2018eyx}}, $a_kH_k=k$, when this mode exit the horizon and $ \mathcal{N}_k$ is defined as the number of e-foldings between the latter and the time inflation ends. Thus, we rewrite \eqref{eqtemp1} as
\begin{eqnarray}
  T_{re}=\left(\frac{43}{11g_{re}}\right)^{\frac{1}{3}}\left(\frac{a_0T_0}{k}\right)H_ke^{- \mathcal{N}_{k}}e^{- \mathcal{N}_{re}}.\label{eqtemp2}
\end{eqnarray}
Inserting \eqref{eqtemp2} in \eqref{eqNre1}, we obtain:
\begin{eqnarray}
   \mathcal{N}_{re}=\frac{4}{3(1+w_{re})}\left[\frac{1}{4}\ln\left(\frac{45}{\pi^2g_{re}}\right)+\ln\left(\frac{V_{end}^{\frac{1}{4}}}{H_k}\right)+\frac{1}{3}\ln\left(\frac{11g_{re}}{43}\right)+\ln\left(\frac{k}{a_0T_0}\right)+ \mathcal{N}_k+ \mathcal{N}_{re}\right],
\end{eqnarray}
or still
\begin{eqnarray}
   \mathcal{N}_{re}=\frac{4}{(1-3w_{re})}\left[-\frac{1}{4}\ln\left(\frac{45}{\pi^2g_{re}}\right)-\ln\left(\frac{V_{end}^{\frac{1}{4}}}{H_k}\right)-\frac{1}{3}\ln\left(\frac{11g_{re}}{43}\right)-\ln\left(\frac{k}{a_0T_0}\right)- \mathcal{N}_k\right].\label{eqNre2}
\end{eqnarray}
Finally, assuming $g_{re}\approx 100$ and using Planck’s pivot of $0.05$Mpc$^{-1}$, one can simplify the expression for $ \mathcal{N}_{re}$ to be 
\begin{eqnarray}
   \mathcal{N}_{re}=\frac{4}{1-3w_{re}}\left[61.6 - \ln\left(\frac{V_{end}^{\frac{1}{4}}}{H_k}\right)- \mathcal{N}_k\right],\label{eqNref}
\end{eqnarray}
where the terms $V_{end}$, $H_k$ and $ \mathcal{N}_k$ depend on the specific inflationary model. The expression \eqref{eqtemp2} turns into
\begin{eqnarray}
  T_{re}=\left[\left(\frac{11g_{re}}{43}\right)^{\frac{1}{3}}\left(\frac{k}{a_0T_0}\right)H_ke^{- \mathcal{N}_k}\left(\frac{45V_{end}}{\pi^2g_{re}}\right)^{-\frac{1}{3(1+w_{re})}}\right]^{\frac{3(1+w_{re})}{3w_{re}-1}}.\label{eqTref}
\end{eqnarray}

The set of equations \eqref{eqNref} and \eqref{eqTref} are used to derive reheating constraints as a function of inflationary model parameters. Note, however, that we need to compute $ \mathcal{N}_k$, $H_k$, and $V_{end}$ for a particular model of interest. The quantity $ \mathcal{N}_k$ can be calculated using the usual definition of e-foldings \eqref{eq:efolds}. The term $V_{end}$ can be obtained by considering $\epsilon=1$ in \eqref{eq:epsilon}. Lastly, the quantity $H_k$, can be derived using the definition of the tensor-to-scalar ratio $r=P_t/P_R$, where $P_t=(2H^2)/(\pi^2M_{Pl}^2)$ and $P_R=A_s$ (at the pivot scale). Then, using $r=16\epsilon$ we get
\begin{eqnarray}
  H_k=\pi M_{Pl}\sqrt{8A_s\epsilon_k}.
\end{eqnarray}
Considering the Arctan potential \eqref{eq:arctan_pot} discussed in the previous section, we can calculate the model dependent parameters of reheating $H_k$ and $V_{end}$, using Eq. \eqref{eq:epsilon}, and $ \mathcal{N}_k$ using Eq.\eqref{eq:ns}. However, to this last one, we had to solve \eqref{eq:ns} numerically in order to obtain $\phi=\phi(\beta,n_s)$, and then use it to calculate $ \mathcal{N}_k= \mathcal{N}_k(n_s)$, $ \mathcal{N}_{re}= \mathcal{N}_{re}(n_s)$ and $T_{re}=T_{re}(n_s)$. 

\begin{figure}[!ht]
 \includegraphics[scale=0.52]{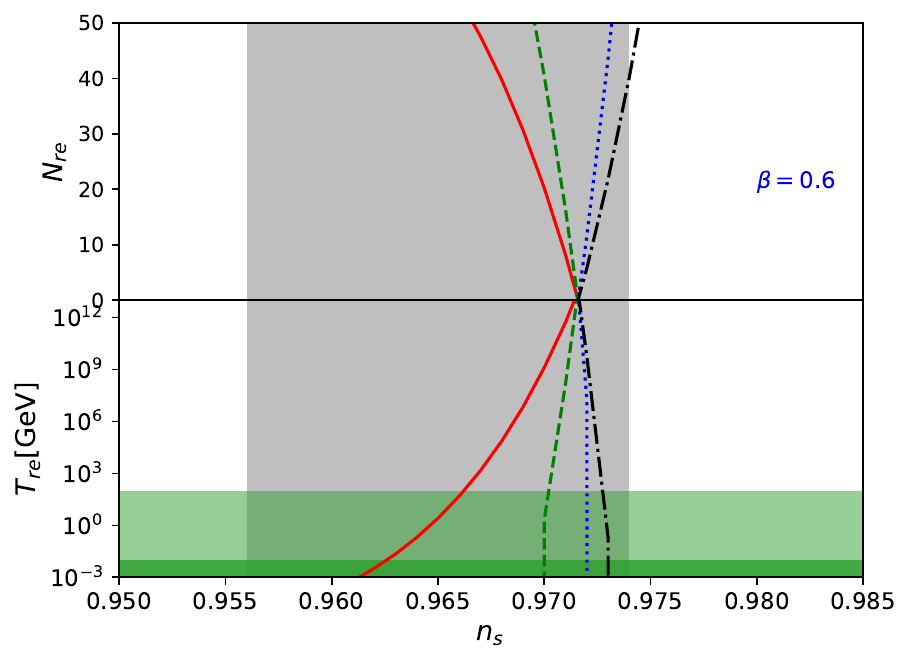}
 \includegraphics[scale=0.52]{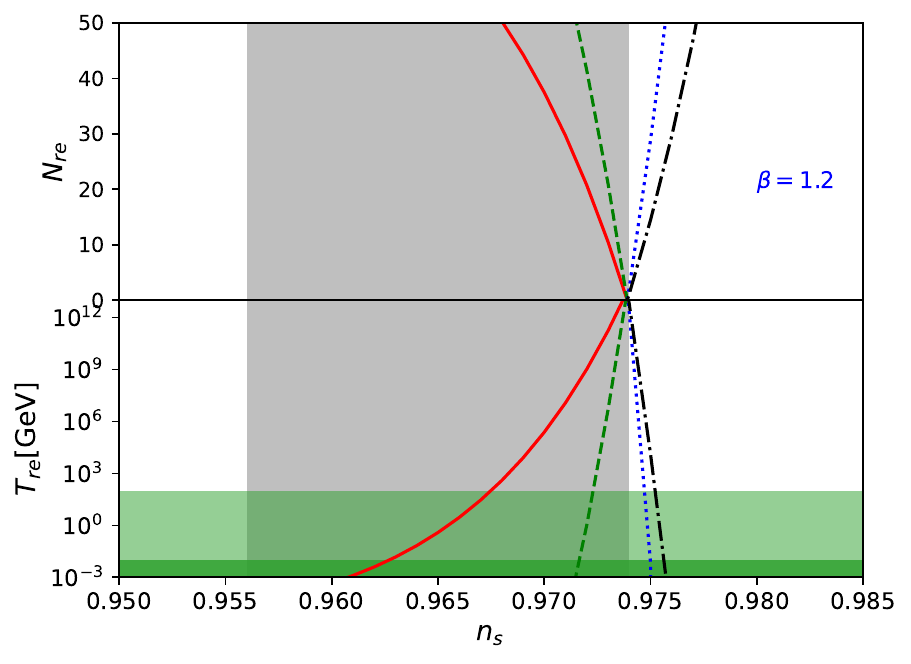}
  \includegraphics[scale=0.52]{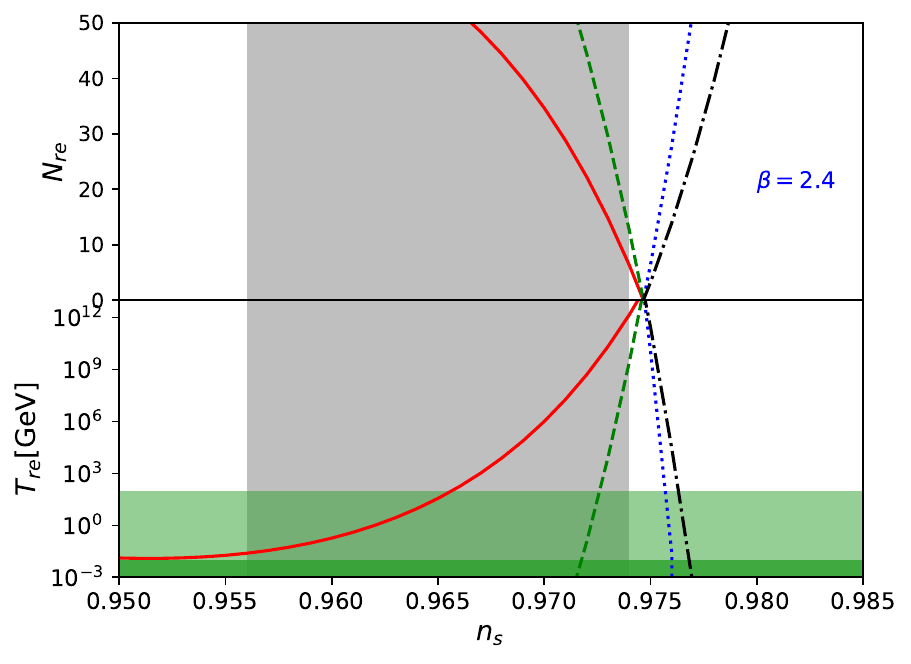}
 \includegraphics[scale=0.52]{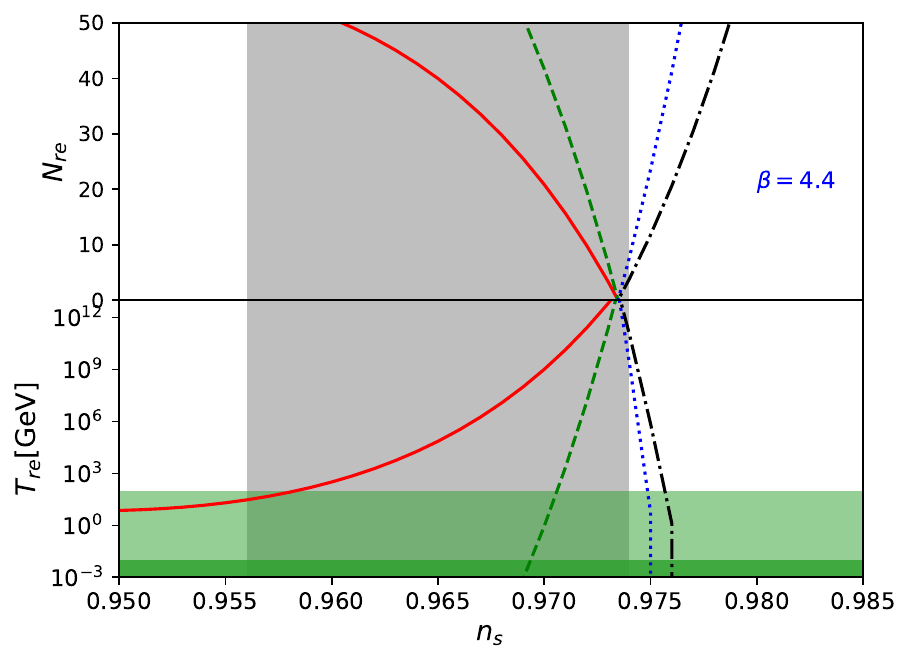}
 \caption{Predictions of $ \mathcal{N}_{re}$ and $T_{re}$, the length of reheating and the temperature at the end
of reheating respectively, for the Arctan potential considering different values of $\beta$. The solid red line
corresponds to $w_{re} = -1/3$, the dashed green line to $w_{re} = 0$, the dotted blue line to $w_{re} = 2/3$, and the dot-dashed black line to $w_{re} = 1$. The grey shaded region corresponds to the $2\sigma$ bounds on $n_s$ from Planck. The light green shaded region is below the electroweak scale, assumed 100 GeV for reference. The dark green region corresponds to the lower bounds of $\mathcal{O}(1 MeV)$ coming from BBN constraints~\cite{Kawasaki:1999na,Kawasaki:2000en}.} \label{fig:NreTre}
\end{figure}

In Fig.~\eqref{fig:NreTre}, we plot the predictions for $ \mathcal{N}_{re}$ and $T_{re}$ for $\beta=0.6,1.2,2.4$ and $4.4$. The best scenario is the one for $\beta=0.6$, where the predictions for all values of the equation of state parameter (different solid lines) relies within the $2\sigma$ interval for $n_s$. The predominant effect in increasing $\beta$ is to shift all the lines towards higher values of $n_s$. However, the case $\beta=2.4$ and $w_{re}>2/3$ (which is out of the $2\sigma$ interval for $n_s$) seems to be an upper limit, from which the increase of $\beta$ brings back the predictions in agreement with the data (see the bottom right panel with $\beta=4.4$). {Note that $\mathcal{N}_{re}$ and $T_{re}$ are directly related with $\mathcal{N}_k(n_s)$, with $n_s$ written in terms of $\eta$, that takes into account the second derivative of the potential. This latter, in turn, is not linear with the increasing of $\beta$ and then could be the origin for this behaviour.}

Note also that all the lines are spread out with the increase of $\beta$, with the most prominent case being $w_{re}=-1/3$. {Moreover, larger values of $\beta$ exhibit a steeper potential and a higher values for its first derivative, as we can see in the left and middle panels of Fig.\ref{fig:pot_derivs}. Thus, if we take into account the definition of the slow-roll parameters \eqref{eq:epsilon} and \eqref{eq:ns}, we find that the higher the value of $V'$, the higher the value of $n_s$ (or the higher is the variation on $n_s$). On the other hand, for small values of $\beta$, we find a plateau potential, with small variations on $V'$ leading to small variations on $n_s$. Therefore, the spread of the lines in Fig.\eqref{fig:NreTre} means that we obtain higher variations in $n_s$ due to the higher variations in the first derivative value of the potential, for higher values of $\beta$.}
{The last two panels show that for all values of $n_s$, the case $w_{re}=-1/3$ obeys very well the BBN constraints for the reheating temperature above the limit of 1 MeV~\cite{Kawasaki:1999na,Kawasaki:2000en}.} Lastly, it is worth mentioning that the limit $ \mathcal{N}_{re}\rightarrow 0$ defines the instantaneous reheating, where all the lines converge and the maximum temperature is achieved.

\begin{figure}[!ht]
\begin{center}
 \includegraphics[scale=0.7]{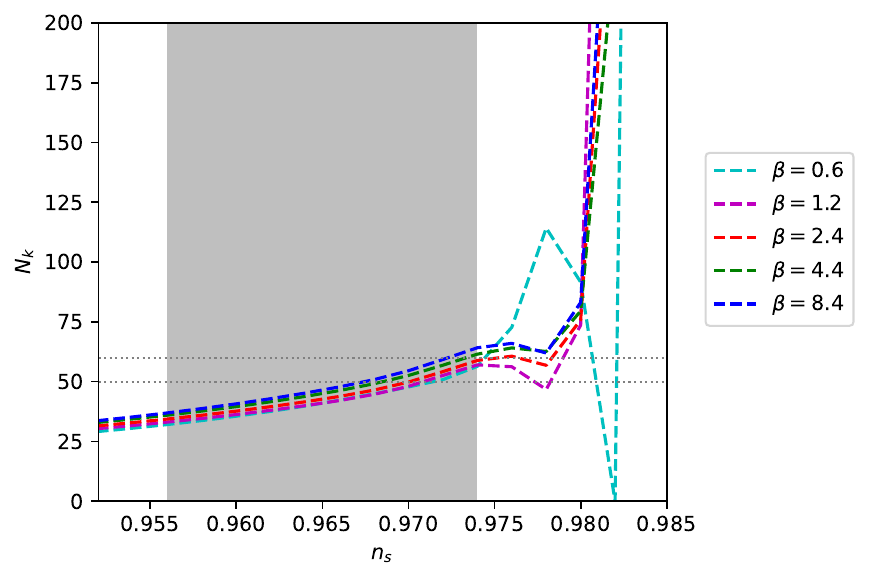}
\caption{Plot of $ \mathcal{N}_k$ in function of $n_s$ for Arctan potential considering different values of $\beta$.}
\end{center}
\label{fig:Nkns}
\end{figure}

Similarly, the result for $ \mathcal{N}_k$ is shown in Fig.\eqref{fig:Nkns}, where we can see that all the values of $\beta$ considered agree with the bounds of $n_s$ (indicated by the grey shaded region). The constraints obtained before for the duration and temperature of reheating are useful here in order to set a minimum value of $ \mathcal{N}_k$, necessary to solve the horizon and flatness problems. We have tried the different scenarios for reheating and all of them exhibit the same behavior displayed in Fig.~\eqref{fig:Nkns}, i.e. with the equation of state parameter being little significant. {We obtained $ \mathcal{N}_k>38.1$ for reheating above the BBN scale for $0.6<\beta<4.4$ and $ \mathcal{N}_k>31.7$ for $\beta>4.4$.} These values are fully consistent with the number of e-folds assumed in the slow-roll analysis, $ \mathcal{N}=50$ and $ \mathcal{N}=60$.

\section{Analysis and Results}\label{analysisandresults}

In order to produce the theoretical predictions for the Arctan model, we adapt the latest version of the Code for Anisotropies in the Microwave Background {\sc CAMB}~\citep{camb} to include the $\beta$ parameter. The version of the Boltzmann solver we used, {\sc Modecode}~\citep{Mortonson:2010er}, is proper to deal with inflationary potentials, by calculating numerically the dynamics, i.e. the Friedmann and Klein-Gordon equations, and the perturbations along with the Fourier components associated with curvature perturbations produced by the fluctuations of the scalar field $\phi$. Then we construct the Primordial Power Spectrum (PPS) which can be translated to the temperature power spectrum of the CMB.

\begin{figure}[!ht]
 \begin{center}
 \includegraphics[scale=0.7]{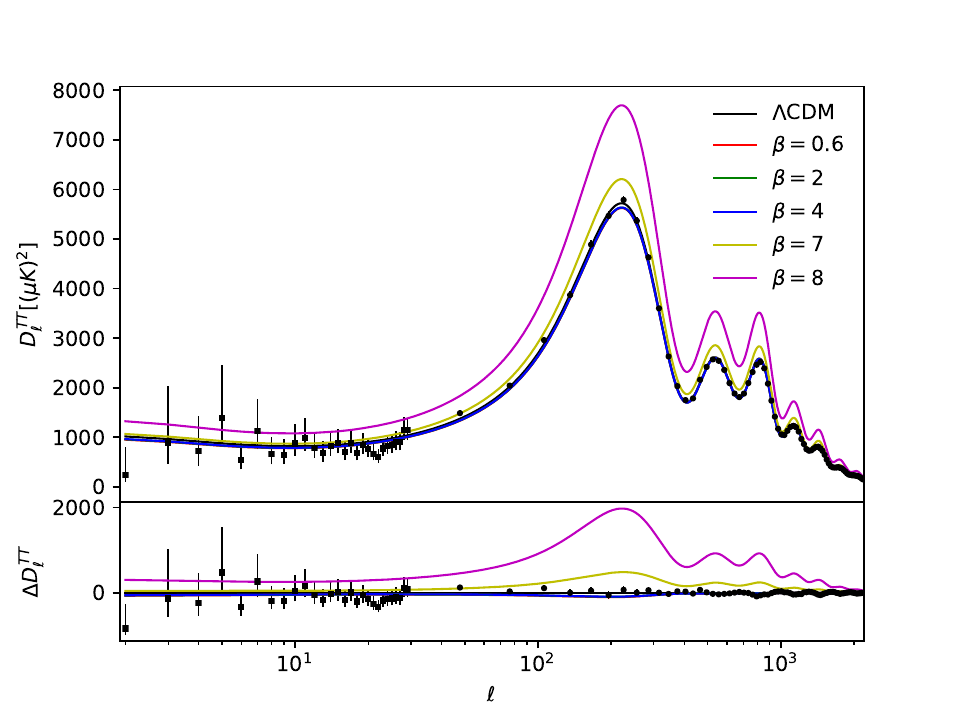}
 \end{center}
\caption{The theoretical predictions for the angular power spectra considering different values of $\beta$.}
\label{fig:arctan_TT_spectrum}
\end{figure}

In general, the method to implement {\sc Modecode} consists in  choosing the potential $V(\phi)$, write its first and second derivatives with respect to the field $\phi$, consider the initial condition to the field, $\phi_{ini}$, and then the code solves the dynamics equations to obtain $H$ and $\phi$. At last, the PPS is obtained through the solutions of the Mukhanov-Sasaki equations~\citep{weinberg2008cosmology}. The PPS can be used to obtain the predictions for the temperature power spectrum of CMB, shown in Fig.\eqref{fig:arctan_TT_spectrum}, considering different values of $\beta$. We observe that the main effect of the $\beta$ parameter is to slightly change the amplitude of the temperature power spectrum, which reinforces as an appropriate range for our analysis the flat prior of $0.6 < \beta < 7$.

%

In order to constrain the cosmological parameters associated with the Arctan model, we perform a Markov Chain Monte Carlo (MCMC) analysis using the latest version of {\sc CosmoMC} code~\citep{cosmomc}.
In addition to the $\beta$ parameter, we vary the usual cosmological variables: the baryon and cold dark matter density, the ratio between the sound horizon and the angular diameter distance at decoupling, and the optical depth: $\left \{\Omega_bh^2~,~\Omega_ch^2~,~\theta~,~\tau\right \}$, respectively. Also, the value of the parameter $\beta$ is chosen according to the considerations made above.
We consider purely adiabatic initial conditions, fix the sum of neutrino masses to $0.06~eV$ and the universe curvature to zero, and also vary the nuisance foregrounds parameters~\citep{Aghanim:2015xee}.
%
In Table \eqref{tab_priors}, we show the  flat priors used in the analysis to all the cosmological parameters.

The data set considered in this analysis comes from the latest Planck (2018) Collaboration release~\citep{Aghanim:2018eyx} and considers the high multipoles Planck temperature data from the 100-,143-, and 217-GHz half-mission T maps, and  the low multipoles data by the joint TT, EE, BB and TE likelihood, where EE and BB are the E- and B-mode CMB polarization power spectrum and TE is the cross-correlation temperature-polarization (hereafter ``PLA18"). 
We also consider an extended data set, combining the CMB data along with i) Baryon Acoustic Oscillations (BAO) from the 6dF Galaxy Survey (6dFGS)~\citep{bao1}, Sloan Digital Sky Survey (SDSS) DR7 Main Galaxy Sample galaxies~\citep{bao2}, BOSSgalaxy samples, LOWZ and CMASS~\citep{bao3} and ii) the tensor amplitude of B-mode polarization, used to constrain the parameters associated with the tensor spectrum coming from 95, 150, and 220 GHz maps, coming from the Keck Array and BICEP2 Collaborations~\citep{bicep21,bicep22} analysis of the BICEP2/Keck field, in combination with Planck high-frequency maps to remove polarized Galactic  dust emission (hereafter ``BKP15'').

%
\begin{table}
\centering
\caption{Priors on the cosmological parameters considered in the analysis.}
{\begin{tabular}{|c|c|}
\hline 
Parameter & Prior Ranges \\ 
\hline
$\Omega_{b}h^{2}$ & $[0.005 : 0.1]$ \\ 
 
$\Omega_{c}h^{2}$ & $[0.001 : 0.99]$ \\ 

$\theta$ & $[0.5 : 10.0]$ \\ 
 
$\tau$ & $[0.01 : 0.8]$ \\ 
 
 
 
$\beta$ & $[0.6 : 7.0]$ \\ 
\hline 
\end{tabular}\label{tab_priors}}
\end{table} 

\begin{figure}[h]
 \begin{center}
 \includegraphics[scale=0.325]{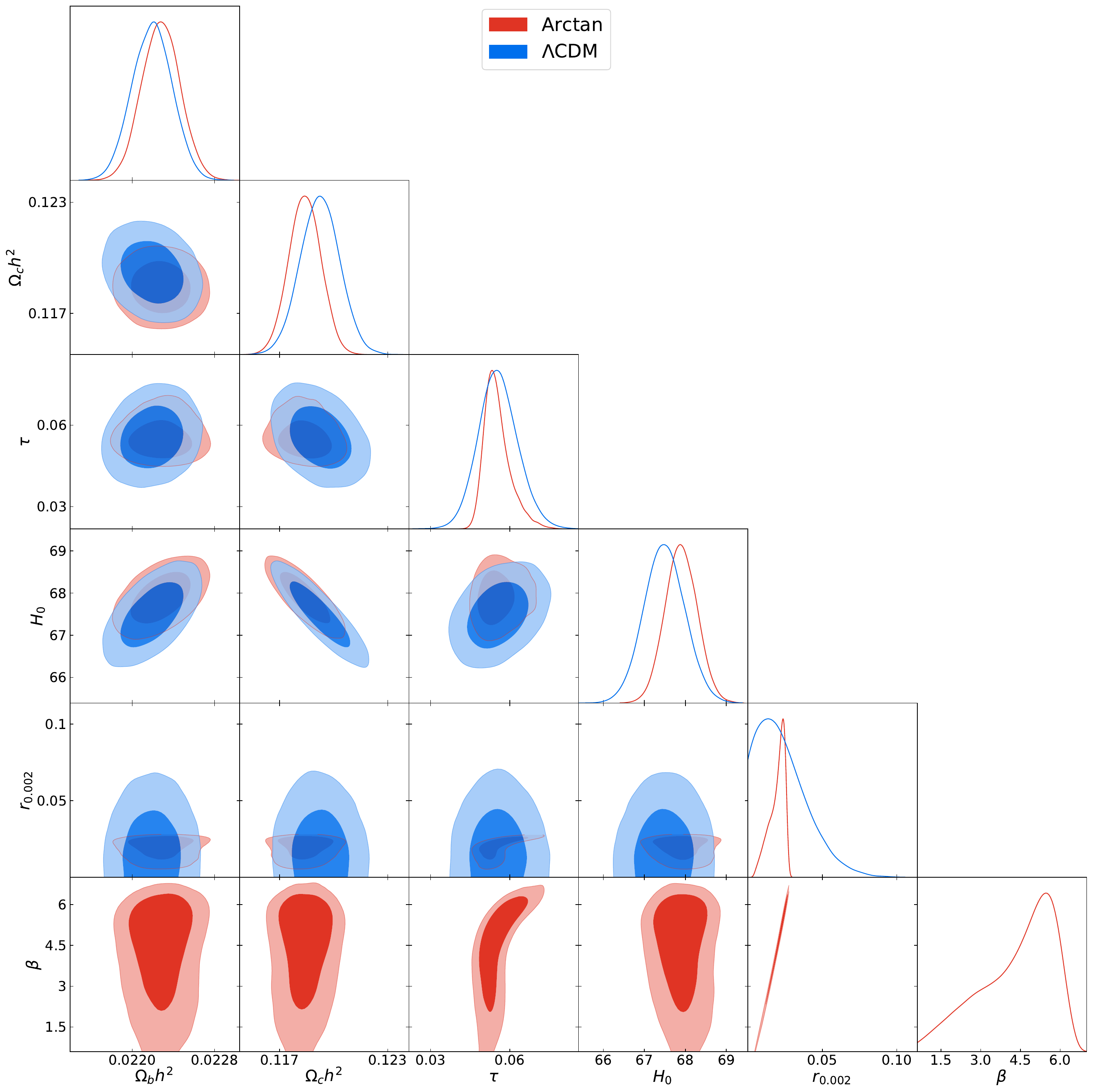}
  \end{center}
\caption{The confidence regions at $68\%$ and $95\%$ and the posterior probability distribution for some primary and derived parameters for the Arctan model.}
\label{fig:arctan_triplot}
\end{figure}

The main results of our analysis are displayed in Table \eqref{tab:Tabel_results_1} and Fig.~\eqref{fig:arctan_triplot}, where we present the main constraints on the cosmological parameters for the Arctan model. 
Note that all the primary cosmological parameters are in good agreement with the $\Lambda$CDM standard model, at least within $2\sigma$~\citep{Aghanim:2018eyx}. We obtain a good constraint on the tensor-to-scalar ratio of $r_{0.002}=0.0196 \pm 0.0052$, while for the $\Lambda$CDM model we find  $r_{0.002}<0.109$~\citep{Aghanim:2018eyx}. Also, this is the first analysis that put a tight restriction on the $\beta$ parameter using observational data, i.e., $\beta=4.37\pm 1.38$ (within $68\%$ confidence level).
{As mentioned previously, higher values of $\beta$ imply on a steeper potential, as seen on Fig.\eqref{fig:pot_derivs}. But $\beta$ is also related to the regime of the inter domain walls distance, as discussed in \cite{Neves:2020lru}. In this way, higher slopes means small inter domain walls distances, which in turn is related to the end of inflation and, consequently, closer to the reheating epoch.}
It is important to note that we analyzed the Arctan potential without making any approximation that can limit the power of constraining the $\beta$ parameter from the slow-roll analysis. In fact, this kind of analysis was performed in \citep{Neves:2020lru} for $\beta = \mathcal{O}(10^{-2})$, where the authors considered an approximation of the type $\phi\gg\beta$.

Lastly, the confidence regions for $68\%$ (C.L.) and $95\%$ (C.L.) and the posterior probability distribution for some primary parameters of the model 
are shown in Fig.~\eqref{fig:arctan_triplot}. 
Notice that the values of the cosmological parameters are well inside the bounds obtained previously for the $\Lambda$CDM model~\citep{Aghanim:2018eyx}. Particularly, our analysis obtains tighter constraints for all parameters, being more prominent the result obtained for the tensor-to-scalar ratio. We also notice a strong correlation between $\beta$ and $r$. 
Finally, we show the best-fit values of the temperature power spectrum for the model, which accommodate the CMB data as good as $\Lambda$CDM cosmology.

\begin{table}
\centering
\caption{{
$68\%$ confidence limits for the cosmological parameters for Arctan model using PLA18+BAO+BK15 data.}
\label{tab:Tabel_results_1}}
\begin{tabular}{|c|c|c|c|}
\hline
{Primary parameters}& 
& {Derived parameters} & 
\\
\hline
$\Omega_b h^2$
& $0.02228 \pm 0.00018$       
&$H_0$
& $67.89 \pm 0.38$ 
\\
$\Omega_{c} h^2$
& $0.1184 \pm 0.0009$ 
&$\Omega_m$
& $0.307 \pm 0.005$
\\
$\theta$
& $1.04107 \pm 0.00041$ 
&$\Omega_{\Lambda}$
& $0.693 \pm 0.005$
\\
$\tau$
& $0.0553 \pm 0.0049$ 
&$n_s$
& $0.9719 \pm 0.0562$
\\
$\beta$
& $4.37 \pm 1.38$
&$r_{0.002}$
& $0.0196 \pm 0.0052$
\\
\hline
\end{tabular}
\end{table} 

%

\begin{figure}[t]
 \begin{center}
 \includegraphics[scale=0.7]{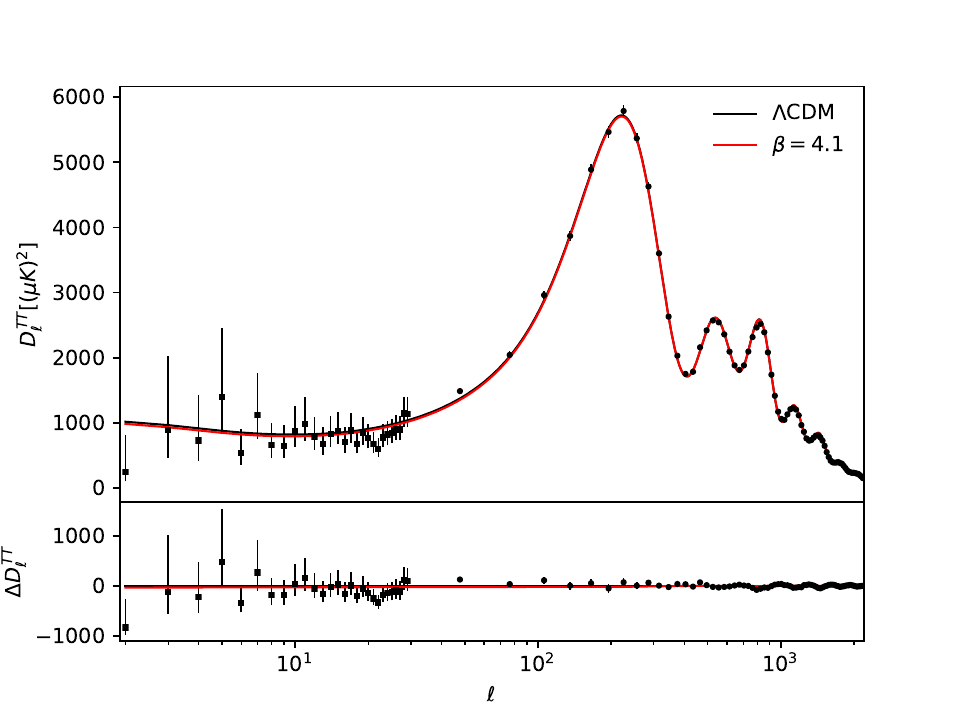}
 \end{center}
\caption{The best-fit angular power spectrum for the Arctan model (blue line) and $\Lambda$CDM model (black dashed line). The data points correspond
to the latest release of Planck (2018) data and the lower panel show the residuals with respect to the reference model ($\Lambda$CDM).}
\label{fig:TT_arctan_bestfit}
\end{figure}
%

\section{Conclusions}\label{conclusions}

In this paper, we analyzed an inflationary model retrieved in a brane cosmology scenario, considering that the period of inflation occurred in a $3$D domain wall immersed in a five-dimensional Minkowski space in the presence of a stack of $N$ parallel domain walls. We then studied the theoretical and observational predictions of an Arctan-type inflationary model and obtained constraints on reheating as well as on its inflationary parameters. 

Regarding the reheating analysis, we showed that the equation-of-state parameter does not influence the predictions of the number of e-folds at the horizon crossing, $\mathcal{N}_k$ (as shown in Fig.\eqref{fig:Nkns}). We found a lower limit of $\mathcal{N}_k>31.8$ for the reheating phase to happen above the BBN scale, which is consistent with the values adopted in the slow-roll analysis when constructing the $n_s-r$ plane. In this case, we found that the range of values $0.6<\beta<6$ ($\mathcal{N}=50$) are in a good agreement with the Planck at least within $2\sigma$ C.L. However, this result contrasts with the analysis performed in \citep{Neves:2020lru}, that obtained values of $\beta$ of order $10^{-2}$ in agreement with the CMB data. Note, however, that the authors of the latter analysis used the approximation $\phi\gg\beta$, while here we analyzed the full range for the Arctan potential.

Finally, we performed the parameter estimation using the latest CMB temperature data combined with BAO and B-mode polarization data. The results of the MCMC analysis for the primary and derived cosmological parameters show an excellent match to the latest cosmological data and with the predictions for the $\Lambda$CDM model (as we can see in Fig.~\eqref{fig:TT_arctan_bestfit}). Particularly, we have obtained the following bound on the tensor-to-scalar ratio $r_{0.002}=0.0196\pm 0.0052$ at $68\%$ C.L., which in principle can be detected by the future CMB experiments that are planned to have sensitivities of order $\Delta r\sim 0.001$~\citep{Ade:2018sbj,Suzuki:2018cuy}. We also obtained good constraints on the $\beta$ parameter that scales the primordial inflationary potential, to be $\beta=4.37\pm 1.38$ ($68\%$ C. L.){, indicating a potential with a high slope or small inter domain walls distance. }
Especially, when considering the bestfit value of $\beta=4.1$, we set the prediction for the inflationary parameters $n_s$ and $r$ right within the $1\sigma$ region allowed by the CMB data and also yield an excellent fit to the temperature power spectrum, as good as the one predicted by the $\Lambda$CDM model (see  Fig.~\eqref{fig:TT_arctan_bestfit}). 

Therefore, the current analysis demonstrates the observational viability of the Arctan-type model and establishes the general theoretical predictions of the model. 

\acknowledgments

R.M.P. Neves is supported by Coordena\c{c}\~{a}o de Aperfei\c{c}oamento de Pessoal de N\'ivel Superior (CAPES).
S.~Santos da Costa thanks the financial support from the Programa de Capacita\c{c}\~ao Institucional (PCI) do Observat\'orio Nacional/MCTI.
F.A. Brito acknowledge support from CNPq (Grant nos. 312104/2018-9). We would like to thanks PRONEX/CNPq/FAPESQ-PB (Grant no. 165/2018), for partial financial support. J.~Alcaniz is supported CNPq (Grants no. 310790/2014-0 and 400471/2014-0) and Funda\c{c}\~ao de Amparo \`a Pesquisa do Estado do Rio de Janeiro FAPERJ (grant no. 233906).
We also thank the authors of the ModeCode (M.~Mortonson, H.
~Peiris and R.~Easther) and CosmoMC (A.~Lewis) codes. Finally, we acknowledge the computational support of the Observat\'orio Nacional Data Center where this work was developed.


\begin{thebibliography}{99}

\bibitem{Aghanim:2015xee}
  N.~Aghanim {\it et al.} [Planck Collaboration],
  Astron.\ Astrophys.\  {\bf 594}, A11 (2016)

\bibitem{Planck} \textrm{Planck Collaboration and P. A. R. Ade \textsl{et al}.,} Astron. Astrophys. \textbf{594}, A20 (2016)

\bibitem{Planckk} \textrm{Planck Collaboration and P. A. R. Ade \textsl{et al}.,} Astron. Astrophys. \textbf{594}, A13 (2016)

\bibitem{ijjas} A. Ijjas, P. J. Steinhardt and A. Loeb, Phys. Lett. \textbf{B 723}, 261 (2013)

\bibitem{Lindee} {A. Linde}, 
[arXiv:1402.0526[hep-th]]

\bibitem{guth14} A. H. Guth, D. I. Kaiser and Y. Nomura, Phys. Lett. \textbf{B 733}, 112 (2014)
 
\bibitem{brandenberger}R. H. Brandenberger, Class. Quant. Grav. \textbf{32}, no. 23, 234002 (2015)

\bibitem{SantosdaCosta:2017ctv} \textrm{S. Santos da Costa, M. Benetti and J. Alcaniz,} 
JCAP {\bf1803}, 004 (2018) 

\bibitem{Santos} \textrm{M. A. Santos, M. Benetti, J. S. Alcaniz, F. A. Brito and R. Silva,} 
JCAP {\bf 1803},  023 (2018) 

\bibitem{bbq-2007} 
  L.~Barosi, F.~A.~Brito and A.~R.~Queiroz,
  JCAP {\bf 0804}, 005 (2008)

\bibitem{Bhattacharya:2020gnk}
S.~Bhattacharya, K.~Dutta, M.~R.~Gangopadhyay, A.~Maharana and K.~Singh,
Phys. Rev. D \textbf{102}, 123531 (2020)


\bibitem{Ohta:2003ie}
N.~Ohta,
Prog. Theor. Phys. \textbf{110}, 269-283 (2003)

\bibitem{gibbons} G.W. Gibbons, 
Int. J. Mod. Phys. A{\bf16}, 822 (2001) 

\bibitem{KKLT-0}  S. Kachru, R. Kallosh, A. Linde and S.P. Trivedi, Phys. Rev. D{\bf68}, 046005 (2003)

\bibitem{townsend} P.K. Townsend and M.N.R. Wohlfarth, Phys. Rev. Lett. {\bf91}, 061302 (2003) 

\bibitem{KKLT} S. Kachru, R. Kallosh, A. Linde, J. Maldacena, L. McAllister and S.P. Trivedi, JCAP {\bf0310}, 013 (2003)

\bibitem{swamp0} H. Ooguri, E. Palti, G. Shiu, C. Vafa, Phys. Lett.B{\bf788} (2019) 180-184

\bibitem{swamp1} W.H. Kinney, S. Vagnozzi, L. Visinelli, Class. Quant. Grav. {\bf36} (2019) 117001

\bibitem{swamp2}A. Mohammadi, T. Golanbari, S. Nasri, K. Saaidi, [arXiv:2006.09489 [gr-qc]]

\bibitem{swamp3} S. Das, Phys. Rev. D{\bf99}, 083510 (2019)

\bibitem{Garg:2018reu}
S.~K.~Garg and C.~Krishnan,
JHEP \textbf{11}, 075 (2019)

\bibitem{Brito} \textrm{F. A. Brito, F. F. Cruz and J. F. N. Oliveira,} Phys. Rev. D\textbf{71}, 083516 (2005)

\bibitem{Neves:2020lru}
R.~M.~P.~Neves, F.~F.~Santos and F.~A.~Brito,
Phys. Lett. B \textbf{810}, 135813 (2020)

\bibitem{Dvali} \textrm{G. Dvali and S.-H. H. Tye,} Phys. Lett. B{\bf450}, 72 (1999)  

\bibitem{M} \textrm{M. Cvetic,} Int. J. Mod. Phys. A\textbf{16}, 819 (2001) 

\bibitem{F} \textrm{F. A. Brito, M. Cvetic and S. -C. Yoon,} Phys. Rev. D\textbf{64}, 064021 (2001)

\bibitem{D} \textrm{D. Bazeia, F. A. Brito and J. R. Nascimento.} Phys. Rev. D\textbf{68}, 085007 (2003)

\bibitem{Bazeia} \textrm{D. Bazeia, F. A. Brito and F. G. Costa,} Phys. Lett. B \textbf{661}, 179 (2008)

\bibitem{Wang:1997cw}
L.~M.~Wang, V.~F.~Mukhanov and P.~J.~Steinhardt,
Phys. Lett. B \textbf{414} (1997), 18-27

\bibitem{Martin:2013nzq}
J.~Martin, C.~Ringeval, R.~Trotta and V.~Vennin,
JCAP \textbf{03} (2014), 039

\bibitem{Martin:2013tda}
J.~Martin, C.~Ringeval and V.~Vennin,
Phys. Dark Univ. \textbf{5-6} (2014), 75-235

\bibitem{liddle}
\textrm{A.~R.~Liddle and D.~H.~Lyth}.
Cambridge university press, 2000.

\bibitem{Lyth_1999}
Lyth, David H., and Antonio Riotto. 
Physics Reports 314.1-2 (1999): 1-146.

\bibitem{Aghanim:2018eyx} 
  N.~Aghanim {\it et al.} [Planck Collaboration],
  arXiv:1807.06209 [astro-ph.CO].


\bibitem{Dai:2014jja}
L.~Dai, M.~Kamionkowski and J.~Wang,
Phys. Rev. Lett. \textbf{113} (2014), 041302


\bibitem{Munoz:2014eqa}
J.~B.~Munoz and M.~Kamionkowski,
Phys. Rev. D \textbf{91} (2015) no.4, 043521

\bibitem{Cook:2015vqa}
J.~L.~Cook, E.~Dimastrogiovanni, D.~A.~Easson and L.~M.~Krauss,
JCAP \textbf{04} (2015), 047

\bibitem{Ueno:2016dim}
Y.~Ueno and K.~Yamamoto,
Phys. Rev. D \textbf{93} (2016) no.8, 083524

\bibitem{Abbott1982}
Abbott L.~F., Farhi E., Wise M.~B., 1982, PhLB, 117, 29. 

\bibitem{Dolgov:1982th}
A.~D.~Dolgov and A.~D.~Linde,
Phys. Lett. B \textbf{116} (1982), 329


\bibitem{Albrecht:1982mp}
A.~Albrecht, P.~J.~Steinhardt, M.~S.~Turner and F.~Wilczek,
Phys. Rev. Lett. \textbf{48} (1982), 1437

\bibitem{Kawasaki:1999na}
M.~Kawasaki, K.~Kohri and N.~Sugiyama,
Phys. Rev. Lett. \textbf{82}, 4168 (1999)

\bibitem{Kawasaki:2000en}
M.~Kawasaki, K.~Kohri and N.~Sugiyama,
Phys. Rev. D \textbf{62}, 023506 (2000)

\bibitem{camb}
A. Lewis, A. Challinor and A. Lasenby, 
Astrophys. J. {\bf 538}, 473 (2000)

\bibitem{Mortonson:2010er}
M.~J.~Mortonson, H.~V.~Peiris and R.~Easther,
Phys. Rev. D \textbf{83}, 043505 (2011)

\bibitem{weinberg2008cosmology}
 S. Weinberg, ``Cosmology", Oxford: OUP OXford (2008).
 
\bibitem{cosmomc} 
A. Lewis and S. Bridle, 
Phys. Rev. D {\bf 66}, 103511 (2002)

\bibitem{bao1}
F.  Beutler, {\it et al.}  
Mon. Not. R. Astron. Soc. {\bf 416}, 3017 (2011)

\bibitem{bao2}
A. J. Ross, {\it et al.} 
Mon. Not. R. Astron. Soc. {\bf 449}, 835 (2015)

\bibitem{bao3}
L. Anderson \textit{et al.}(BOSS Collaboration), Mon. Not. R.Astron. Soc.441, 24 (2014)

\bibitem{bicep21}
P. A. R. Ade {\it et al.} (BICEP2 and Planck Collaborations), Phys. Rev. Lett. {\bf 114}, 101301 (2015)

\bibitem{bicep22}
P. A. R. Ade {\it et al.} (BICEP2 and Keck Array Collaborations), Phys. Rev. Lett. {\bf 116}, 031302 (2016)

\bibitem{Ade:2018sbj} 
  P.~Ade {\it et al.} [Simons Observatory Collaboration],
  JCAP {\bf 1902}, 056 (2019)

\bibitem{Suzuki:2018cuy} 
  A.~Suzuki {\it et al.},
  J.\ Low.\ Temp.\ Phys.\  {\bf 193}, no. 5-6, 1048 (2018)
  
%
  















\end{thebibliography}
\end{document}